\documentclass[]{emulateapj}
\usepackage{natbib}
\usepackage{amsmath}
\usepackage{graphicx}
\usepackage{color}
\shortauthors{K. Inayoshi et al.}
\shorttitle{stellar pulsation v.s. maser periodicity}

\newcommand{\msunyr}{{M}_{\sun}~{\rm yr}^{-1}}
\newcommand{\mdot}{\dot{M}_{\ast}}
\newcommand{\msun}{{M}_{\sun}}
\newcommand{\rsun}{{R}_{\sun}}
\newcommand{\lsun}{{L}_{\sun}}

\begin{document}

\title{direct diagnostics of forming massive stars: \\
stellar pulsation and periodic variability of maser sources}
\author{Kohei Inayoshi\altaffilmark{1},
Koichiro Sugiyama\altaffilmark{2},
Takashi Hosokawa\altaffilmark{3},
Kazuhito Motogi\altaffilmark{4},
Kei E. I. Tanaka\altaffilmark{5,1}
}

\altaffiltext{1}{Department of Physics, Graduate School of Science, Kyoto University, 
Kyoto 606-8502, Japan; inayoshi@tap.scphys.kyoto-u.ac.jp}
\altaffiltext{2}{Graduate School of Science and Engineering, Yamaguchi University, 1677-1
Yoshida, Yamaguchi, Yamaguchi 753-8512, Japan; koichiro@yamaguchi-u.ac.jp}
\altaffiltext{3}{Department of Physics, University of Tokyo, 
Tokyo 113-0033, Japan; takashi.hosokawa@phys.s.u-tokyo.ac.jp}
\altaffiltext{4}{The Research Institute for Time Studies, Yamaguchi
University, Yoshida 1677-1, Yamaguchi, Yamaguchi, 753-8511, Japan}
\altaffiltext{5}{Astronomical Institute, Tohoku University, Miyagi 980-8578, Japan}

\begin{abstract}
The 6.7 GHz methanol maser emission, a tracer of forming massive stars, 
sometimes shows enigmatic periodic flux variations over several
$10-100$ days.
In this Letter, we propose that this periodic variations could be
explained by the pulsation of massive protostars growing under rapid
mass accretion with rates of $\mdot \gtrsim 10^{-3}~\msunyr$. 
Our stellar evolution calculations predict that the massive protostars
have very large radius exceeding $100~R_\sun$ at maximum, and 
we here study the pulsational stability of such the bloated protostars
by way of the linear stability analysis.
We show that the protostar becomes pulsationally unstable with
various periods of several $10-100$ days, depending on different 
accretion rates.
With the fact that the stellar luminosity when the star is pulsationally
unstable also depends on the accretion rate, we derive the 
period-luminosity relation 
$\log (L/\lsun) = 4.62 + 0.98\log(P/100~{\rm day})$, which is testable
with future observations.
Our models further show that the radius and
mass of the pulsating massive protostar should also depend on the
period. It would be possible to infer such protostellar properties 
and the accretion rate with the observed period.
Measuring the maser periods enables a direct diagnosis
of the structure of accreting massive protostars, which are 
deeply embedded in dense gas and inaccessible with other observations.
\end{abstract}


%%%%%%%%%%%%%%%%%
%        Sec.1   Introduction         %
%%%%%%%%%%%%%%%%%
\section{Introduction}

Massive stars have significant impacts on the interstellar medium
through various feedback processes such as supernova explosions, stellar
winds, and ultraviolet radiation.
These feedback processes would be also important for
shaping the stellar initial mass function because stars are mostly
formed in clusters including massive stars (e.g., Lada \& Lada 2003).
However, our understanding of the massive star formation is still
limited with observational difficulties, one of which arises from
the fact that forming massive stars are deeply embedded in obscuring 
dense gas.

%--------------------------------------------------------------------------%

Observing maser emission is one of the possible methods to see
the vicinity ($< 10^3$~AU) of forming massive stars 
because of the high brightness.
It is known that the 6.7 GHz methanol maser emission can be thought to be 
often associated with circumstellar disks around forming massive stars
(e.g., Norris et al. 1993; Bartkiewicz et al. 2009; Sanna et al. 2010).
Some of the methanol masers show periodic flux variations over $\gtrsim 10$ days
(e.g., Goedhart et al. 2004, 2009).
Since the 6.7 GHz methanol masers are radiatively pumped by infrared 
emission of warm dusts ($\sim 150$ K; Cragg et al. 2005), 
the periodicity could reflect the luminosity variation of nearby
forming massive stars or accretion disks.

%----------------------------------------------------------------------------%

Several authors have proposed different explanations for this 
periodic flux variations, e.g., colliding-wind binary 
(van der Walt 2011), and periodic accretion onto binary systems
(Araya et al. 2010).
However, these binary-based models do not explain why the periodic
variations shorter than $10$ days have not been found yet, despite the
fact that a number of OB star eclipsing binaries have the periods 
of $< 10$ days (Harries et al. 2003; Hilditch et al. 2005)

%-----------------------------------------------------------------------%

In this Letter, we propose an alternative picture that the periodic
variability of the maser emission can be 
due to the pulsation of protostars growing via
rapid mass accretion with $\mdot \gtrsim 10^{-3}~\msunyr$,
which is expected in the massive star formation
(e.g., Osorio et al. 1999; McKee \& Tan 2003;
Zhang et al. 2005; Beltr{\'a}n et al. 2006; Krumholz et al. 2009).
Our previous work shows that, by numerically modeling the stellar
evolution, the massive protostar should have the large radius exceeding
$100~\rsun$ with such the high accretion rate (Hosokawa \& Omukai
2009; Hosokawa et al. 2010, hereafter HO09 and HYO10
respectively).
We here examine the pulsational stability of the bloated massive
protostars with the linear stability analysis, and 
show that the observed periodicity of $\sim 10 - 100$ days 
can be well explained with the stellar pulsation.
Our models predict that, with the observed period of the pulsation,
we could infer the radius, mass, and luminosity of the protostars 
as well as the accretion rate onto the protostar.
Measuring the maser periods would thus make a direct diagnosis of
the central small-scale ($\lesssim$ a few AU) structure of forming massive
stars, which is beyond observations in the optical and infrared bands.

%%%%%%%%%%%%%%%%%%%%%%%
%        Sec.2   The pulsational instability       %
%           of accreting massive stars              %
%%%%%%%%%%%%%%%%%%%%%%%
\section{Pulsational instability of accreting massive stars}

\begin{figure}
\begin{center}
\includegraphics[height=58mm,width=75mm]{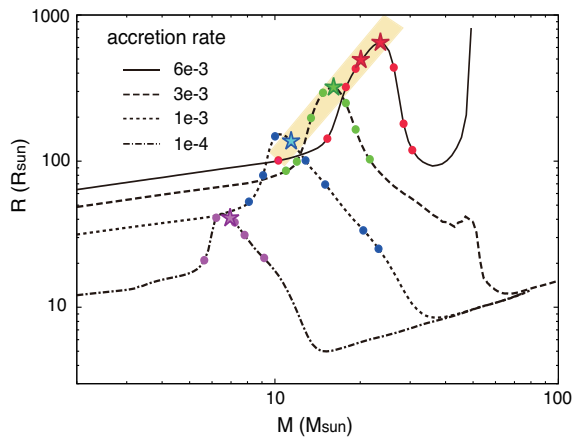}
\end{center}
\caption{Evolution of the protostellar radius with various accretion rates
$\mdot=10^{-4}$, $10^{-3}$, $3\times 10^{-3}$, and $6\times 10^{-3}~\msunyr$
(taken from HO09).
Symbols indicate the stellar models we conduct the linear 
stability analysis. The circles and stars represent the 
pulsationally stable and unstable models.
The shaded layer denotes the {\it instability strip}, 
where the protostar becomes unstable.
}
\label{fig:history}
\end{figure}

We study the pulsational stability of massive protostars
growing at constant accretion rates of $\mdot \geq 10^{-4}~\msunyr$.
We here adopt the protostellar models with spherical accretion taken from HO09.
In Section.~3, we will discuss that the effects of the accretion geometry (disk accretion) 
does not change the qualitative results from the spherical accretion case.
Figure~\ref{fig:history} shows the evolution of the stellar radius
as the stellar mass increases with different accretion rates.
The outline of the evolution is briefly summarized as follows
(see HO09 for the details).
Initially, the stellar radius gradually increases with increasing
the stellar mass.
After this stage, e.g., $M_\ast \gtrsim 10~M_\sun$ for 
$\mdot = 10^{-3}~\msunyr$, the protostar contracts by losing 
its energy via radiation (Kelvin-Helmholtz or KH contraction).
The stellar central temperature increases during this contraction
stage and finally reaches $10^7$~K. The hydrogen burning is 
ignited, and the protostar reaches the zero-age main sequence 
(ZAMS, except with $6 \times 10^{-3}~\msunyr$).
After this point, e.g., $M_* \simeq 40~M_\sun$ for 
$\mdot = 10^{-3}~\msunyr$, the stellar radius increases again
as the stellar mass increases.
We here note that the maximum stellar radius during this evolution 
is larger with the higher accretion rate.
This is because the accreting gas has the higher specific entropy
with the more rapid mass accretion, which leads to the higher
average entropy in the stellar interior (also see HO09).
As a result, the maximum stellar radius exceeds
$100~R_\odot$ with the high accretion rates 
$\mdot \gtrsim 10^{-3}~\msunyr$.

%%%%%%%%%  FIg.1  %%%%%%%%%
%%%%%%%%%%%%%%%%%%%%%%

%-----------------------------------------------------------------------%

We apply the linear stability analysis to the above protostellar models
(see e.g., Cox 1980; Unno et al. 1989; Inayoshi et al. 2013 
for the details).
We here consider radial (spherical) perturbations with 
the time dependence of $e^{i\sigma t}$, 
where $\sigma =\sigma _{\rm R} + i \sigma _{\rm I}$ is the eigen 
frequency, $\sigma _{\rm R}$ is the frequency of the pulsation,
and $|\sigma _{\rm I}|$ is the growth or damping rate of the
perturbation.
The protostar is pulsationally stable (unstable) with the positive 
(negative) $\sigma_{\rm I}$.
If unstable, the perturbation grows until reaching the non-linear 
regime, where the pulsation energy is dissipated by shock waves near 
the stellar surface.
The dissipated energy is partly converted into the kinetic energy 
of periodic outflows (e.g., Appenzeller 1970; Yoon \& Cantiello 2010).

%------------------------------------------------------------------------%

Symbols in Figure~\ref{fig:history} represent the stellar models 
for which we conduct the linear stability analysis.
Our calculations show that the protostar becomes pulsationally unstable 
only when the stellar radius maximally expands at a given accretion
rate.
This instability is caused by the $\kappa$ mechanism in the He$^+$ 
ionization layer, where the radiative energy flux is blocked and 
converted into the pulsation energy (e.g., Cox 1980; Unno et al. 1989).
In the KH contraction stage, the stellar surface temperature increases 
and the He$^+$ ionization layer disappears.
The protostar is consequently stabilized and the pulsation ceases.
Although the protostar would be also unstable
with the lower accretion rates ($\la 10^{-3}~\msunyr$), 
the growth time is much longer 
than the duration for which the star is in the instability strip
($\sigma_{\rm I}^{-1}\gg M_{\ast}/\mdot \sim 10^4$ yr).
The perturbation does not grow enough to cause the stellar
pulsation in this case.
The instability strip thus does not extend for 
$M_* \lesssim 10~\msun$, where the star becomes unstable
with the lower accretion rate
(see Figure \ref{fig:history}).

%%%%%%%%%  Fig.2  %%%%%%%%%
\begin{figure}
\begin{center}
\includegraphics[height=60mm,width=75mm]{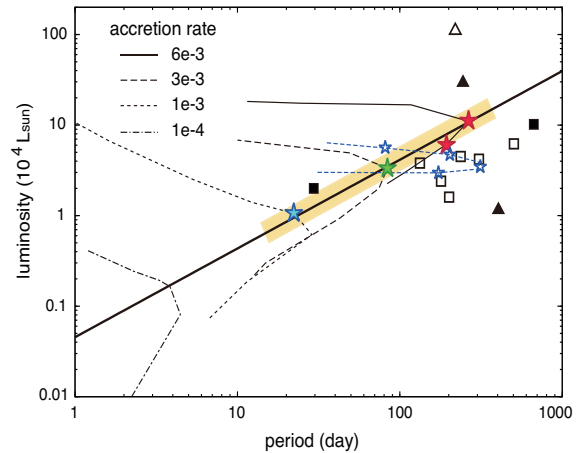}
\end{center}
\caption{
The period-luminosity (P-L) relation of forming massive protostars.
Thin black (blue) curves show the evolutionary tracks with the 
spherical (disk) accretion at the rates of
$\mdot =10^{-4}$, $10^{-3}$, $3\times 10^{-3}$, and $6\times 10^{-3}~\msunyr$.
The star symbols on the tracks denote the pulsationally 
unstable models. 
The eigen frequencies are plotted for the stable stellar models.
The shaded layer shows the instability strip
as in Figure~\ref{fig:history}.
The solid line represents the P-L relation given by 
equation (\ref{eq:P-L}), which fits the unstable models.
The filled and open symbols represent the observed sources, 
whose distances are measured with the trigonometric parallax
and kinematics. The triangles indicate that the sources are 
associated with ultra/hypercompact H$_{\rm II}$ regions
(Walsh et al. 1998, Reid et al. 2009a, Urquhart et al. 2007).
}
\label{fig:P-L}
\end{figure}
%%%%%%%%%%%%%%%%%%%%%%

%%%%%%%%%%%%%%%%%%%%%%%%%%%%
%           Sec.3   The period-luminosity relation              %
%%%%%%%%%%%%%%%%%%%%%%%%%%%%
\section{Period-Luminosity relation}

Figure~\ref{fig:P-L} shows the evolution of the pulsation period
and stellar luminosity in the examined cases.
In each case, the stellar luminosity monotonically increases
as the stellar mass increases. 
The pulsation period increases with the stellar mass
in the early expansion phase, and decreases in the later
KH contraction phase.
The protostar becomes pulsationally unstable around the turning
point of the period, which corresponds to that 
of the stellar radius as seen in Figure~\ref{fig:history}.

%------------------------------------------------------------------------%

Figure~\ref{fig:P-L} also present the observed sources with 
periodic flux variations in the 6.7~GHz methanol masers,
whose parameters are summarized in Table~\ref{tab:tab1}.
The luminosities of these sources are estimated with 
the far-infrared data of Infrared Astronomical Satellite (IRAS)  
following \cite{1986A&A...169..281C} and \cite{2012ApJ...753...51G}. 
We see that the pulsation periods of the unstable
models are several $10-100$ days, the same order of the observed
periods of the maser sources.
Although the pulsation period is shorter than $10$ days with
the low accretion rates $\mdot \la 10^{-3}~\msunyr$, 
the instability strip does not cover such cases
as explained above.
This explains well why the maser sources which have such the
short periodic variability have not been observed.

%----------------------------------------------------------------------%

Our calculations predict that both the period and luminosity
of the pulsationally unstable protostars increase with the accretion rate.
We thus predict a positive correlation between the 
maser periodicity and the stellar luminosity.
We fit the unstable models by a single power law and obtain the
period-luminosity (P-L) relation (thick line in Fig.~\ref{fig:P-L})
\begin{equation}
\log (L/\lsun) = 4.62 + 0.98\log(P/100~{\rm days}),
\label{eq:P-L}
\end{equation}
which is similar to that of the Cepheids ($L\propto P^{4/3}$), 
which is well-used as a cosmic distance ladder 
(e.g., Tammann et al. 2003; Sandage et al. 2004).
The above P-L relation and the instability 
strip can explain 
the distribution of the observed parameters 
in the periodic methanol masers 
within errors of an order of magnitude.
We show that, in Appendix~\ref{sec:app}, 
a similar P-L relation is analytically derived.

%-------------------------------------------------------------------%

The instability strip shown in Figures~\ref{fig:history} and 
\ref{fig:P-L} is not broad. The duration for which a protostar
spends in the strip, i.e., the prospective lifetime of each periodic
maser source is only $\sim 10^3$ years.
This actually matches the rarity expected with observations.
The number of 6.7 GHz methanol masers observed ever is $\sim 900$
(e.g., Pestalozzi et al. 2005; Caswell et al. 2011; Green et al. 2012),
and 56 masers of that have been monitored for more than a year  
at intervals shorter than a week or month
\citep{
2007IAUS..242...97G,2009MNRAS.398..995G,2010ApJ...717L.133A,
2011A&A...531L...3S}.
About $20$~\% of such the sources show the characteristic 
periodic variabilities (Table 1).
Figure~\ref{fig:P-L} might suggest that our models are more
applicable to the sources with the shorter periods, e.g., 
$P \la 0.5$~year, which is only $\simeq 5$~\% of the monitored
sources.
Since the appearance time of the methanol masers during 
massive star formation is thought to be $\simeq 3\times 10^{4}$ yr 
\citep{2005MNRAS.360..153V},
the lifetime of the periodic maser sources should be
as short as expected with our calculations.

%%%%%%%%%  Tab.1  %%%%%%%%%
\begin{table}[t]
\begin{center}
\caption{Methanol maser sources with periodic variability}
\label{tab:tab1}
\begin{tabular}{lccccc}
\hline\hline
Source & $P_{\mathrm{met}}$ & $D_{\mathrm{src}}$ & $L_{\mathrm{iras}}$ & \multicolumn{2}{c}{Ref.} \\ \cline{5-6}
 &  &  &  & $P$ & $D$ \\
(Galactic~name)   & [day] & [kpc] & [$10^{4}~{\lsun}$] &  &  \\ \hline
009.621$+$0.196 &                         244  & \hspace{1.5mm}5.2   & \hspace{1.5mm}30.0  & 1 & 6 \\
012.681$-$0.182 &                         307  & \hspace{1.5mm}4.5   & \hspace{3mm}4.2      & 2 & 7 \\
012.889$+$0.489 & \hspace{4mm}29.5      & \hspace{3mm}2.34   & \hspace{3mm}2.0      & 3 & 8 \\
022.357$+$0.066 & \hspace{2.5mm}179.2  & \hspace{1.5mm}4.6  & \hspace{3mm}2.4      & 4 & 9 \\
037.550$+$0.200 &                         237  & \hspace{1.5mm}5.0   & \hspace{3mm}4.5      & 5 & 9 \\
188.946$+$0.886 &                         404  & \hspace{3mm}2.10    & \hspace{3mm}1.2      & 1 & 10 \\
196.454$-$1.677 &                         668  & \hspace{3mm}5.28    & \hspace{1.5mm}10.2  & 2 & 11 \\
328.237$-$0.547 &                         220  &                       12.0  &                     117.2  & 1 & 9 \\
331.132$-$0.244 &                         504  & \hspace{1.5mm}4.7   & \hspace{3mm}6.2      & 1 & 7 \\
338.935$-$0.062 &                         133  & \hspace{1.5mm}2.9   & \hspace{3mm}3.8      & 1 & 9 \\
339.622$-$0.121 &                         201  & \hspace{1.5mm}2.6   & \hspace{3mm}1.6      & 1 & 9 \\
\hline
\end{tabular}
\tablecomments{Col.~1: name of the source, 
Col.~2: variability period, Col.~3: distance,
Col.~4: luminosity estimated with the IRAS data,
Col.~5 and 6: references of the periods (P) and distances (D).}
\tablerefs{\scriptsize (1)~\cite{2007IAUS..242...97G};
(2)~\cite{2004MNRAS.355..553G};
(3)~\cite{2009MNRAS.398..995G};
(4)~\cite{2011A&A...531L...3S};
(5)~\cite{2010ApJ...717L.133A};
(6)~\cite{2009ApJ...706..464S};
(7)~Near kinematic distance\tablenote{\scriptsize
Assumed on a flat rotation curve with a Galactic circular rotation
of the Sun, of 246~km~s$^{-1}$ \citep{2009ApJ...704.1704B},
and a solar distance, of 8.4~kpc
(e.g., Reid et al. 2009b).
};
(8)~\cite{2011ApJ...733...25X};
(9)~\cite{2011MNRAS.417.2500G};
(10)~\cite{2009ApJ...693..397R};
(11)~\cite{2007PASJ...59..889H}.
}
\end{center}
\end{table}
%%%%%%%%%%%%%%%%%%%%%%%%%

An uncertainty of the P-L relation arises from possible
variations of our stellar evolution tracks.
While HO09 study the protostellar evolution with the spherical
accretion, for instance, HYO10 present that the evolution
slightly changes with different accretion geometries, e.g., 
accretion via a geometrically thin disk.
%-------------------------------------------------------------%
\footnote{
While the ``disk accretion'' mentioned here assumes 
the lowest entropy of the accreting gas, the rapid mass accretion 
could enhance the entropy (e.g., Hosokawa et al. 2011). 
In this case, the stellar evolution with the spherical accretion would 
be more realistic even if gas accretes through the disk (see HYO10).
}
%---------------------------------------------------------------%
We also perform the stability analysis of case MD3-D-b0.1
in HYO10, whose model has the largest radius among 
the cases for $\mdot =10^{-3}~\msunyr$.
Figure~\ref{fig:P-L} shows that the protostar also becomes 
pulsationally unstable in this case, and that the periods
are 10 times longer than those with the spherical accretion
at the same rate. 
This effect might explain the observed sources which have
the longer periods than predicted by the P-L relation (\ref{eq:P-L}).

%%%%%%%%%%%%%%%%%%%%%%%%%%%%
%               Sec.4   Conclusion and Discussion             %
%%%%%%%%%%%%%%%%%%%%%%%%%%%%

\section{Conclusion and Discussion}

%-------------------------------------------------------------%

In this Letter, we have shown that the pulsation of 
massive protostars could explain the periodic 
variability of the 6.7 GHz maser sources.
Our linear stability analysis clearly shows that 
a rapidly accreting ($\mdot \ga10^{-3}~\msunyr$)
massive protostar becomes pulsationally unstable 
when the star is bloated before reaching the ZAMS.
Typical periods of the pulsation are several $10-100$ days, 
which explain the observed periodicity well.
The period depends on the adopted accretion rate, getting
longer with the higher rate.
On the other hand, the protostar with the lower $\mdot$ 
becomes unstable but does not produce the periodicity, 
which could also explain why the periodic variability 
shorter than 10 days has not been observed.
Our stability analysis predicts the period-luminosity (P-L) 
relation for the pulsationally unstable protostars,
$\log (L/\lsun) = 4.62 + 0.98\log(P/100~{\rm day})$.

%-----------------------------------------------------------------------%

\begin{figure}
\begin{center}
\includegraphics[height=60mm,width=75mm]{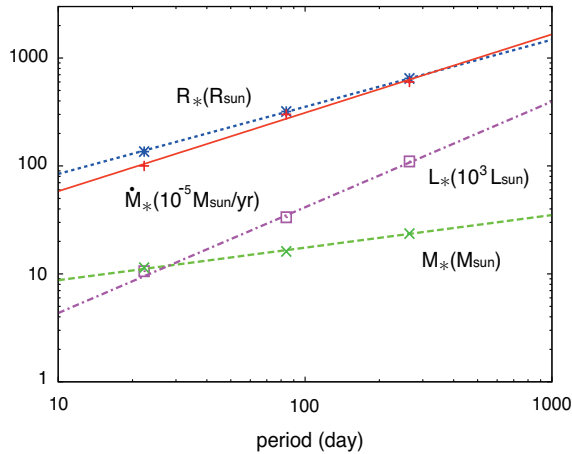}
\end{center}
\caption{
Dependences of the protostellar properties (the stellar
radius $R_{\ast}$, luminosity $L_{\ast}$, mass $M_{\ast}$,
and accretion rate $\dot{M_\ast}$)
on the period in the stellar-pulsation models. 
Each line fits the results of the stability analysis 
(symbols) by a single power-law function 
(see equations \ref{eq:P-L} - \ref{eq:MdotP}).
}
\label{fig:prediction}
\end{figure}

Moreover, our stellar evolution models predict that other stellar
properties such as the mass, radius, and accretion rate as well
as the luminosity should depend on the pulsation period. 
We show this in Figure~\ref{fig:prediction}.
Single power-law functions which fit the results are, 
\begin{equation}
M_{\ast} = 17.5~\msun \left(\frac{P}{100~{\rm days}}\right)^{0.30},
\label{eq:MP}
\end{equation}
\begin{equation}
R_{\ast} = 350~\rsun \left(\frac{P}{100~{\rm days}}\right)^{0.62},
\label{eq:RP}
\end{equation}
\begin{equation}
\dot{M}_{\ast} = 3.1\times 10^{-3}~\msunyr 
\left(\frac{P}{100~{\rm days}}\right)^{0.73}.
\label{eq:MdotP}
\end{equation}
If the P-L relation (\ref{eq:P-L}) is confirmed by 
further observations, we can infer the above quantities
with equations (\ref{eq:MP}) - (\ref{eq:MdotP}) 
only from the maser period.
As seen in Figure \ref{fig:P-L}, the periods of the observed maser
sources are longer than 10~days.
Our stellar-pulsation models predict that, 
in particular to explain the periods longer than $100$ days, 
very rapid mass accretion with $\mdot \ga 3 \times 10^{-3}~\msunyr$
is required.
In this way, we can make a direct diagnosis of the small-scale
structure of accreting massive protostars and their vicinities
($\la$ a few AU), which are difficult to see in the optical 
and infrared bands, by measuring the maser periods.

%%%%%%%%%  Fig.4  %%%%%%%%%
%%%%%%%%%%%%%%%%%%%%%%

%---------------------------------------------------------------%

The stellar pulsation examined above should cause the
periodic variation of the stellar luminosity $L_\ast$.
Although it is still uncertain where the maser emission 
is excited around protostars, some observations suggest that 
the maser sources are located on circumstellar disks 
at $r \sim 10^3$ AU (e.g., Bartkiewicz et al. 2009).
The temperature of the circumstellar disk irradiated
by the stellar radiation obeys $T_{\rm d} (r) \simeq 120~{\rm K}~
(L_\ast/3\times10^4~\lsun)^{1/4}(r/10^3~{\rm AU})^{-1/2}$,
where $r$ is the distance from the protostar.
The variation of the stellar luminosity $\delta L_\ast$ 
changes the dust temperature following  
$\delta T_{\rm d} / T_{\rm d} \simeq 
\delta L_\ast / 4 L_\ast$.
Since the response of the dust temperature to the irradiating
flux variation is so rapid ($\sim $ a few sec), 
$\delta T_{\rm d}$ should synchronize with 
$\delta L_\ast$.
As evaluating $\delta L_\ast$ is beyond our linear analysis, 
we here consider the flux variation of Mira variables, 
which are also excited by the $\kappa$-mechanism. 
Interestingly, SiO maser sources associated with some Mira
variables are likely pumped by the stellar radiation, and
show periodic variations which synchronize with
the stellar pulsation (Pardo et al. 2004). 
With the typical flux variations of $\simeq 3$ magnitudes of
Mira-type stars (Samus et al. 2009), 
the amplitude of the variation is estimated as 
$\delta T_{\rm d} / T_{\rm d} \simeq 0.3$, 
which results in $\delta T_{\rm d} \simeq 40$~K around
$T_{\rm d} \simeq 120$~K.
Cragg et al. (2005) show that the variations of the dust temperature 
$\delta T_{\rm d} \ga 20$~K could produce the observed amplitudes 
of the maser variability ($\sim 10-100$ Jy).
Therefore, the stellar pulsation could produce the flux variations 
of the observed 6.7 GHz methanol masers.

%--------------------------------------------------------------------%

In the above, we have supposed that the stellar radiation 
reaches the irradiated disk surface without significant time delays,
i.e., the light path from the star to the disk surface 
is optically thin. 
This should be valid with the disk accretion, where the density 
above the disk quickly decreases as the gas falls onto the disk. 
If outflow is launched from the disk, however, the disk wind
would enhance the density above the disk and increase the optical
depth. 
According to a disk wind model by Zhang, Tan \& McKee (2013), 
the optical depth when the protostar is pulsationally unstable
is estimated as 
$\simeq 5~(\mdot/10^{-3}~\msunyr)^{0.27}
(\bar \kappa/3~{\rm cm}^2~{\rm g}^{-1})$,
where $\bar \kappa$ is the mean opacity of dusts.
The diffusion time with this optical depth is $\simeq 30$ days, 
which is shorter than the typical periods of the maser sources.
This effect would smear out only variabilities shorter than 
a few $10$ days.
Nonetheless, the launching mechanism of the disk wind and the 
detailed density structure above the disk are still highly uncertain.
Some periodic maser sources which have $P \ga 100$~days
also show rapid fluctuations of light curves over a few 10 days
(G009.621, G022.357, and G338.935).
The above estimate of the diffusion time do not explain 
such the rapid changes. 
Although the variations of these sources might be explained 
with different models rather than ours (also see below), 
it would be also possible that the disk wind is weaker
and the diffusion time is shorter than evaluated above.
Simultaneous monitoring of the maser sources in the infrared band 
will verify the relation between the periodic variations of the stellar luminosity 
and the maser variability.

%-----------------------------------------------------------------%

Finally, we compare our pulsation model to alternative models 
for explaining the periodic variability of the 6.7 GHz 
methanol maser sources.
For instance, van der Walt (2011) proposes that radiation from
shocked gas in the colliding binary wind explains
both the periodicity and shapes of the light curves
of some maser sources.
This model supposes a particular condition, the binary
system surrounded by an ultracompact H$_{\rm II}$ region
($\sim 10^3$ AU), and explains the periodic maser sources
for which ultracompact H$_{\rm II}$ regions have been observed
(triangles in Fig~\ref{fig:P-L}). 
On the other hand, a number of the sources
presented in Figure~\ref{fig:P-L} do not show associating 
H$_{\rm II}$ regions at least at 5 and 8 GHz. 
Our models could explain these sources, as the protostar 
becomes pulsationally unstable when the stellar radius is very large
and effective temperature is only $T_{\rm eff} \simeq 5000$~K, with
which the UV luminosity is too low to create a detectable H$_{\rm II}$
region.
This suggests that measuring the UV luminosities of 
massive protostars would be a key for discriminating the models.
Further observations at the higher frequencies 
($>$ a few tens GHz) would detect hypercomact
H$_{\rm II}$ regions, which are too small and optically
thick to be detected at the lower frequencies. 
Such the observations with the higher spatial resolution
(e.g., with ALMA) are sure to advance our understanding
of the enigmatic periodic variability of the maser sources.

{\acknowledgements 
We would like to thank Kazuyuki Omukai, Munetake Momose, Yoshinori Yonekura, 
Kenta Fujisawa, Mareki Honma, Yichen Zhang, Takashi Nakamura, Yudai Suwa, 
and Daisuke Nakauchi for their fruitful discussions.
This work is supported in part by the Grants-in-Aid by the Ministry 
of Education, Culture, and Science of Japan (23$\cdot $838 KI; 24$\cdot$6525 KM)
and by the project of Yamaguchi University entitled 
``The East-Asian VLBI Network and the Circulation 
of Matter in the Universe".\\
}

\appendix
\section{Analytic derivation of the P-L relation}
\label{sec:app}

In this Appendix, we analytically derive a P-L relation similar
to equation (\ref{eq:P-L}) with considering the protostellar
evolution.
Stahler et al. (1986, hereafter SPS86) show that the
evolution of an accreting protostar is well understood with 
the balance between following two timescales: 
the accretion time scale $t_{\rm acc}={M_\ast}/{\mdot}$,
and the Kelvin-Helmholtz timescale 
$t_{\rm KH}={GM_\ast ^2}/{L_\ast R_\ast}$.
As seen in Figure \ref{fig:history}, the protostar gradually expands
in the early stage with $t_{\rm acc} < t_{\rm KH}$, and turns to 
contract in the later stage with $t_{\rm acc} > t_{\rm KH}$.
The turnaround of the radius occurs when $t_{\rm acc} \simeq t_{\rm KH}$
(also see HO09 and HYO10).
According to our linear stability analysis, the protostar 
becomes unstable at the epoch of the maximum radius, 
i.e., $t_{\rm acc} \simeq t_{\rm KH}$ 
(see star symbols in Figure \ref{fig:history}).

%-------------------------------------------------------------------------%

SPS86 also show that the stellar expansion in the early evolutionary 
stage is well described as $R_\ast \propto {M_\ast}^{0.27}{\mdot}^{0.41}$.
With Kramers' law $\kappa \propto \rho T^{-7/2}$, which approximates 
the dependences of opacity in the stellar interior in this early
stage, the stellar luminosity generally obeys 
$L_{\ast}\propto {M_\ast}^{11/2}{R_\ast}^{-1/2}$ 
(e.g., Cox \& Giuli 1968).
Using these relations, we can express the luminosity of the unstable protostar 
(i.e., $t_{\rm acc} \simeq t_{\rm KH}$) 
as a function of the pulsation period $P~(\propto R_{\ast}^{3/2}M_{\ast}^{1/2})$
and obtain the period-luminosity (P-L) relation $L_\ast \propto P^{1.2}$.


\begin{thebibliography}{99}

\bibitem[Appenzeller(1970)]{1970A&A.....5..355A} Appenzeller, I.\ 1970, A\&A, 5, 355 
\bibitem[Araya et al.(2010)]{2010ApJ...717L.133A} Araya, E.~D., Hofner, P., 
Goss, W.~M., et al.\ 2010, \apjl, 717, L133
\bibitem[Bartkiewicz et al.(2009)]{2009A&A...502..155B} Bartkiewicz, A., Szymczak, M., van Langevelde, H.~J., Richards, A.~M.~S., \& Pihlstr{\"o}m, Y.~M.\ 2009, A\&A, 502, 155
\bibitem[Beltr{\'a}n et al.(2006)]{2006Natur.443..427B} Beltr{\'a}n, M.~T., 
Cesaroni, R., Codella, C., et al.\ 2006, Nature, 443, 427 
\bibitem[Bovy et al.(2009)]{2009ApJ...704.1704B} Bovy, J., Hogg, D.~W., 
\& Rix, H.-W.\ 2009, \apj, 704, 1704 
\bibitem[Casoli et al.(1986)]{1986A&A...169..281C} Casoli, F., Combes, F.,
Dupraz, C., Gerin, M., \& Boulanger, F.\ 1986, \aap, 169, 281 
\bibitem[Caswell et al.(2011)]{2011MNRAS.417.1964C} Caswell, J.~L., Fuller, 
G.~A., Green, J.~A., et al.\ 2011, \mnras, 417, 1964 
\bibitem[Cox 
\& Giuli(1968)]{1968QB801.C65......} Cox, J.~P., \& Giuli, R.~T.\ 1968, Principles of Stellar Structure, New York, Gordon and Breach
\bibitem[Cox(1980)]{1980tsp..book.....C} Cox, J.~P.\ 1980, Theory of Stellar Pulsation,
Princeton University Press, Princeton, NJ
\bibitem[Cragg et al.(2005)]{2005MNRAS.360..533C} Cragg, D.~M., Sobolev, 
A.~M., \& Godfrey, P.~D.\ 2005, MNRAS, 360, 533 
\bibitem[Goedhart et al.(2004)]{2004MNRAS.355..553G} Goedhart, S., Gaylard, 
M.~J., \& van der Walt, D.~J.\ 2004, MNRAS, 355, 553 
\bibitem[Goedhart et al.(2007)]{2007IAUS..242...97G} Goedhart, S., Gaylard, 
M.~J., \& van der Walt, D.~J.\ 2007, IAU Symposium, 242, 97 
\bibitem[Goedhart et al.(2009)]{2009MNRAS.398..995G} Goedhart, S., Langa, 
M.~C., Gaylard, M.~J., \& van der Walt, D.~J.\ 2009, MNRAS, 398, 995 
\bibitem[Green \& McClure-Griffiths(2011)]{2011MNRAS.417.2500G}
Green, J.~A., \& McClure-Griffiths, N.~M.\ 2011, \mnras, 417, 2500
\bibitem[Green et al.(2012)]{2012MNRAS.420.3108G} Green, J.~A., Caswell, 
J.~L., Fuller, G.~A., et al.\ 2012, \mnras, 420, 3108 
\bibitem[Guzm{\'a}n et al.(2012)]{2012ApJ...753...51G} Guzm{\'a}n, A.~E., 
Garay, G., Brooks, K.~J., \& Voronkov, M.~A.\ 2012, \apj, 753, 51 
\bibitem[Harries et al.(2003)]{2003MNRAS.339..157H} Harries, T.~J., 
Hilditch, R.~W., \& Howarth, I.~D.\ 2003, MNRAS, 339, 157 
\bibitem[Hilditch et al.(2005)]{2005MNRAS.357..304H} Hilditch, R.~W., 
Howarth, I.~D., \& Harries, T.~J.\ 2005, MNRAS, 357, 304 
\bibitem[Honma et al.(2007)]{2007PASJ...59..889H} Honma, M., Bushimata, T., 
Choi, Y.~K., et al.\ 2007, \pasj, 59, 889  
\bibitem[Hosokawa 
\& Omukai(2009)]{2009ApJ...691..823H} Hosokawa, T., \& Omukai, K.\ 2009, ApJ, 691, 823 (HO09)
\bibitem[Hosokawa et al.(2010)]{2010ApJ...721..478H} Hosokawa, T., Yorke, 
H.~W., \& Omukai, K.\ 2010, ApJ, 721, 478 (HYO10)
\bibitem[Hosokawa et al.(2011)]{2011ApJ...738..140H} Hosokawa, T., Offner, 
S.~S.~R., \& Krumholz, M.~R.\ 2011, \apj, 738, 140
\bibitem[Inayoshi et al.(2013)]{2013MNRAS.tmp.1079I} Inayoshi, K., 
Hosokawa, T., \& Omukai, K.\ 2013, MNRAS, 1079 
\bibitem[Krumholz et al.(2009)]{2009Sci...323..754K} Krumholz, M.~R., 
Klein, R.~I., McKee, C.~F., Offner, S.~S.~R., 
\& Cunningham, A.~J.\ 2009, Science, 323, 754 
\bibitem[Lada 
\& Lada(2003)]{2003ARA&A..41...57L} Lada, C.~J., \& Lada, E.~A.\ 2003, \araa, 41, 57 
\bibitem[McKee 
\& Tan(2003)]{2003ApJ...585..850M} McKee, C.~F., \& Tan, J.~C.\ 2003, ApJ, 585, 850 
\bibitem[Norris et al.(1993)]{1993ApJ...412..222N} Norris, R.~P., Whiteoak,
J.~B., Caswell, J.~L., Wieringa, M.~H.,
\& Gough, R.~G.\ 1993, ApJ, 412, 222
\bibitem[Osorio et al.(1999)]{1999ApJ...525..808O} Osorio, M., Lizano, S., 
\& D'Alessio, P.\ 1999, ApJ, 525, 808 
\bibitem[Pardo et 
al.(2004)]{2004A&A...424..145P} Pardo, J.~R., Alcolea, J., Bujarrabal, V., et al.\ 2004, A\&A, 424, 145 
\bibitem[Pestalozzi et al.(2005)]{2005A&A...432..737P}
Pestalozzi, M.~R., Minier, V., \& Booth, R.~S.\ 2005, \aap, 432, 737 
\bibitem[Reid et al.(2009a)]{2009ApJ...693..397R} Reid, M.~J., Menten, 
K.~M., Brunthaler, A., et al.\ 2009a, \apj, 693, 397 
\bibitem[Reid et al.(2009b)]{2009ApJ...700..137R} Reid, M.~J., Menten, 
K.~M., Zheng, X.~W., et al.\ 2009b, \apj, 700, 137 
\bibitem[Samus et al.(2009)]{2009yCat....102025S} Samus, N.~N., Durlevich,
O.~V., \& et al.\ 2009, VizieR Online Data Catalog, 1, 2025
\bibitem[Sandage et 
al.(2004)]{2004A&A...424...43S} Sandage, A., Tammann, G.~A., \& Reindl, B.\ 2004, A\&A, 424, 43 
\bibitem[Sanna et al.(2009)]{2009ApJ...706..464S} Sanna, A., Reid, M.~J., 
Moscadelli, L., et al.\ 2009, \apj, 706, 464 
\bibitem[Sanna et
al.(2010)]{2010A&A...517A..78S} Sanna, A., Moscadelli, L., Cesaroni, R., et al.\ 2010, A\&A, 517, A78
\bibitem[Stahler et al.(1986)]{1986ApJ...302..590S} Stahler, S.~W., Palla, 
F., \& Salpeter, E.~E.\ 1986, ApJ, 302, 590 (SPS86)
\bibitem[Szymczak et al.(2011)]{2011A&A...531L...3S} Szymczak, M., Wolak, P.,
Bartkiewicz, A., \& van Langevelde, H.~J.\ 2011, \aap, 531, L3 
\bibitem[Tammann et 
al.(2003)]{2003A&A...404..423T} Tammann, G.~A., Sandage, A., \& Reindl, B.\ 2003, A\&A, 404, 423 
\bibitem[Unno et al.(1989)]{1989nos..book.....U} Unno, W., Osaki, Y., Ando, 
H., Saio, H., 
\& Shibahashi, H.\ 1989, Nonradial oscillations of stars, 2nd ed. University of Tokyo Press, Tokyo  
\bibitem[Urquhart et al.(2007)]{2007A&A...461...11U} Urquhart, J.~S.,
Busfield, A.~L., Hoare, M.~G., et al.\ 2007, A\&A, 461, 11
\bibitem[van der Walt(2005)]{2005MNRAS.360..153V} van der Walt, J.\ 2005, 
\mnras, 360, 153 
\bibitem[van der Walt(2011)]{2011AJ....141..152V} van der Walt, D.~J.\ 
2011, AJ, 141, 152 
\bibitem[Walsh et al.(1998)]{1998MNRAS.301..640W} Walsh, A.~J., Burton,
M.~G., Hyland, A.~R., \& Robinson, G.\ 1998, MNRAS, 301, 640
\bibitem[Yoon 
\& Cantiello(2010)]{2010ApJ...717L..62Y} Yoon, S.-C., \& Cantiello, M.\ 2010, ApJL, 717, L62 
\bibitem[Xu et al.(2011)]{2011ApJ...733...25X} Xu, Y., Moscadelli, L., 
Reid, M.~J., et al.\ 2011, \apj, 733, 25 
\bibitem[Zhang et al.(2005)]{2005ApJ...625..864Z} Zhang, Q., Hunter, T.~R., 
Brand, J., et al.\ 2005, ApJ, 625, 864 
\bibitem[Zhang et al.(2013)]{2013ApJ...766...86Z} Zhang, Y., Tan, J.~C., 
\& McKee, C.~F.\ 2013, ApJ, 766, 86 



\end{thebibliography}
\end{document}